\documentclass{aastex63}

\usepackage{color}

\received{}
\revised{}
\accepted{}
\submitjournal{ApJL}

\shorttitle{ALMA observations of dense cores in Taurus}
\shortauthors{Fujishiro et al.}

\begin{document}

\newcommand{\ntdp}{N$_2$D$^+$}
\newcommand{\jnco}{$^{12}$CO}
\newcommand{\jsco}{$^{13}$CO}
\newcommand{\cjho}{C$^{18}$O}

\title{A low-velocity bipolar outflow from a deeply embedded object in Taurus revealed by the Atacama Compact Array}

\correspondingauthor{Kazuki Tokuda}
\email{tokuda@p.s.osakafu-u.ac.jp}

\author{Kakeru Fujishiro}
\affiliation{Department of Physics, Nagoya University, Chikusa-ku, Nagoya 464-8602, Japan}

\author[0000-0002-2062-1600]{Kazuki Tokuda}
\affiliation{Department of Physical Science, Graduate School of Science, Osaka Prefecture University, 1-1 Gakuen-cho, Naka-ku, Sakai, Osaka 599-8531, Japan}
\affiliation{National Astronomical Observatory of Japan, National Institutes of Natural Science, 2-21-1 Osawa, Mitaka, Tokyo 181-8588, Japan}

\author{Kengo Tachihara}
\affiliation{Department of Physics, Nagoya University, Chikusa-ku, Nagoya 464-8602, Japan}

\author{Tatsuyuki Takashima}
\affiliation{Department of Physical Science, Graduate School of Science, Osaka Prefecture University, 1-1 Gakuen-cho, Naka-ku, Sakai, Osaka 599-8531, Japan}

\author{Yasuo Fukui}
\affiliation{Department of Physics, Nagoya University, Chikusa-ku, Nagoya 464-8602, Japan}
\affiliation{Institute for Advanced Research, Nagoya University, Furo-cho, Chikusa-ku, Nagoya 464-8601, Japan}

\author{Sarolta Zahorecz}
\affiliation{Department of Physical Science, Graduate School of Science, Osaka Prefecture University, 1-1 Gakuen-cho, Naka-ku, Sakai, Osaka 599-8531, Japan}
\affiliation{National Astronomical Observatory of Japan, National Institutes of Natural Science, 2-21-1 Osawa, Mitaka, Tokyo 181-8588, Japan}

\author{Kazuya Saigo}
\affiliation{National Astronomical Observatory of Japan, National Institutes of Natural Science, 2-21-1 Osawa, Mitaka, Tokyo 181-8588, Japan}

\author{Tomoaki Matsumoto}
\affiliation{Faculty of Sustainability Studies, Hosei University, Fujimi, Chiyoda-ku, Tokyo 102-8160, Japan}

\author{Kengo Tomida}
\affiliation{Astronomical Institute, Tohoku University, 6-3, Aramaki, Aoba-ku, Sendai, Miyagi 980-8578, Japan}

\author{Masahiro N. Machida}
\affiliation{Department of Earth and Planetary Sciences, Faculty of Sciences, Kyushu University, Nishi-ku, Fukuoka 819-0395, Japan}

\author{Shu-ichiro Inutsuka}
\affiliation{Department of Physics, Nagoya University, Chikusa-ku, Nagoya 464-8602, Japan}

\author{Philippe Andr\'e}
\affiliation{Laboratoire d'Astrophysique (AIM), CEA, CNRS, Universit\'e Paris-Saclay, Universit\'e Paris Diderot, Sorbonne Paris Cit\'e, F-91191 Gif-sur-Yvette, France}

\author{Akiko Kawamura}
\affiliation{National Astronomical Observatory of Japan, National Institutes of Natural Science, 2-21-1 Osawa, Mitaka, Tokyo 181-8588, Japan}

\author[0000-0001-7826-3837]{Toshikazu Onishi}
\affiliation{Department of Physical Science, Graduate School of Science, Osaka Prefecture University, 1-1 Gakuen-cho, Naka-ku, Sakai, Osaka 599-8531, Japan}

\begin{abstract}

The first hydrostatic core, the first quasi-hydrostatic object formed during the star formation process, is still the observational missing link between the prestellar and protostellar phases, mainly due to its short lifetime. Although we have not established a clear method to identify this rare object, recent theoretical studies predict that the first core has millimeter continuum emission and low-velocity outflow with a wide opening angle. An extensive continuum/outflow survey toward a large number of $``$starless$"$ cores in nearby star-forming regions works as a pathfinder.
We observed 32 prestellar cores in Taurus with an average density of $\gtrsim$10$^5$\,cm$^{-3}$ in 1.3\,mm continuum and molecular lines using the Atacama Large Millimeter/submillimeter Array--Atacama Compact Array (ALMA--ACA) stand-alone mode. Among the targets, MC35-mm centered at one of the densest $``$starless$"$ cores in Taurus has blueshifted/redshifted wings in the $^{12}$CO\,(2--1) line, indicating that there is deeply embedded object driving molecular outflow. The observed velocities and sizes of the possible outflow lobes are 2--4\,km\,s$^{-1}$, and $\sim$2\,$\times$\,10$^3$\,au, respectively, and the dynamical time is calculated to be $\sim$10$^3$\,yr. In addition to this, the core is one of the strongest N$_2$D$^{+}$\,(3--2) emitters in our sample.  
All of the observed signatures do not conflict with any of the theoretical predictions about the first hydrostatic core so far, and thus MC35-mm is unique as the only first-core candidate in the Taurus molecular cloud.

\end{abstract}


\section{Introduction}\label{sec:intro}
Understanding of the low-mass star formation process has been intensively studied from decades ago, mainly by theoretical work at the beginning \citep[e.g.,][]{Lar1969,Shu1977,Shu1987,Inu2012}.
They explained that the fragmentation and condensation of molecular clouds result in forming dense cores, which undergo gravitational collapse to form stars. According to the theoretical studies, dense cores eventually harbor the first protostellar cores, the first quasi-hydrostatic object during the star formation process (hereafter, the first core; e.g., \citealt{Lar1969,Mas1998,Tom2013}, ), which provide the initial condition of star formation. 
Recent magnetohydrodynamic (MHD) simulations demonstrated that the first core with a size of 1--100\,au is formed when the central density exceeds $\sim$10$^{10}$\,cm$^{-3}$ via the gravitational collapse. The first core is suggested to have a low-velocity (1--10\,km\,s$^{-1}$) molecular outflow with a wide opening angle \citep{Mac2008}, which is qualitatively different from the collimated jet driven by a mature protostar.
However, it is difficult to identify such an object observationally because the first core has a short lifetime, 10$^3$--10$^4$\,yr, depending on the physical condition of the parental core \citep{Tom2010}, and does not show bright infrared emission. Although some candidates of the first core were already reported in the past decade \citep[e.g.,][]{Che2010,Che2012,Eno2010,Pin2011,Pez2012,Hir2014}, the first-core phase is not fully explored observationally and it is supposed to still be the missing link between the isothermal and adiabatic collapse (i.e., prestellar and protostellar core). 
To search for candidates of the first core, it is essential to perform a survey-type observation toward a large number of starless cores. 
According to the early dense core survey with an average density of $\gtrsim$10$^5$\,cm$^{-3}$ \citep{Oni2002}, the lifetime of the starless phase is $\sim$4\,$\times$\,10$^5$\,yr (see also \citealt{War2007}). The simple calculation tells us that only one out of a few $\times$\,10--100 cores harbors the first core(s). 

The Taurus molecular cloud complex \citep[e.g.,][]{Ken2008} is one of the best-studied low-mass star-forming regions, and there is a dense core catalog \citep{Oni2002}, which was sampled by an unbiased large-scale molecular gas survey. We carried out an Atacama Large Millimeter/submillimeter Array (ALMA) survey, FRagmentation and Evolution of dense cores Judged by ALMA (FREJA), toward the dense cores in Taurus using the Atacama Compact Array (ACA). \citet[hereafter, Paper~I]{Tok2020} described the observation settings and early results obtained by the 1.3\,mm continuum emission. In this Letter, we explain the detailed results of a dense core L1535-NE/MC35, harboring a possible candidate of the first core.

\section{Observations and descriptions of the target object, L1535-NE/MC35} \label{sec:obs}
As the detailed observational properties of this project are given in Paper~I, and thus we summarize the data qualities of the target briefly. The angular resolution is 6\farcs8\,$\times$\,6\farcs0, corresponding to 850\,au\,$\times$\,730\,au at a distance of 126.6\,pc \citep{Galli18}. The frequency settings contain the 1.3\,mm continuum emission and molecular lines of $^{12}$CO, $^{13}$CO, C$^{18}$O (2--1), and N$_2$D$^+$(3--2). The rms sensitivities of the continuum and lines are $\sim$0.4\,mJy\,beam$^{-1}$ and $\sim$0.06\,mJy\,beam$^{-1}$ ($\sim$0.05\,K) at a velocity resolution of $\sim$0.1\,km\,s$^{-1}$, respectively. 
We use the combined 7\,m + Total Power (TP) array data to recover the extended emission of $^{12}$CO  (Sect.\,\ref{result:outflow}),  while for \ntdp\, we use the 7\,m array data alone to be compared with the compact 1.3\,mm source (Sect.\,\ref{result:N2DP}). 

\begin{figure}[htbp]
\includegraphics[width=\linewidth]{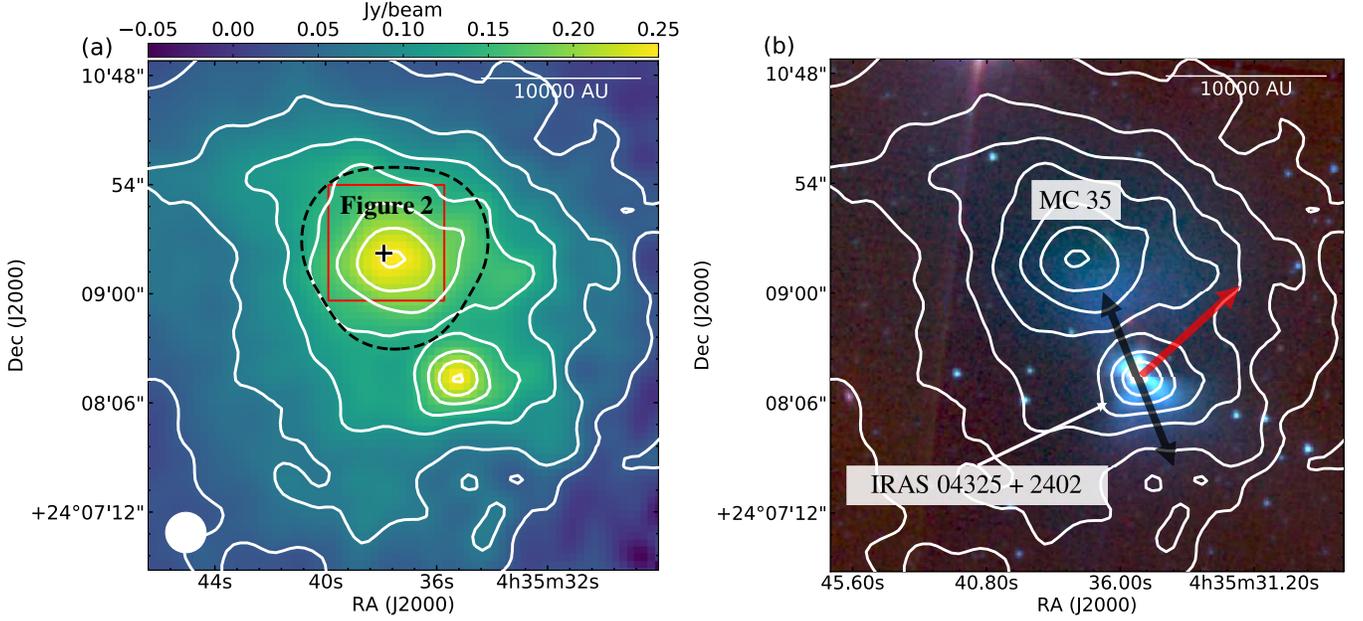}
\caption{(a) The color-scale image and white contour show the 1.2\,mm dust continuum intensity toward L1535-NE/MC35 and IRAS\,04325+2402 taken by IRAM/MAMBO-2 \citep{Kau2008}. 
The dashed contour shows the observation field with the 7\,m array. The black cross denotes the position of 1.3\,mm peak observed with the 7\,m array (see the text in Sect.\,\ref{result:outflow}). 
(b) White contours are the same as panel (a). Color is the R(8\,$\mu$m)–G(4.5\,$\mu$m)–B(3.6\,$\mu$m) image with {\rm Spitzer}. The black and red arrows are the directions of the scattered-light nebula and the redshifted parsec-scale CO outflow (see the text in Sect.\,\ref{firstcore}). 
}
\label{fig:mc35}
\end{figure}

The target object, L1535-NE (also know as MC35; see \citealt{Oni2002}), is located in the Barnard18 region in Taurus. \cite{Hog2000} found an intensity enhancement in the submillimeter continuum emission, observed by JCMT/SCUBA at 850\,$\mu$m and with IRAM\,30\,m/MAMBO at 1.2\,mm by \cite{Mot2001} next to the previously know protostar IRAS\,04325+2402, which is a Class\,I source with multiplicity \citep{Har1999}. The nondetection of infrared point sources at the center of the dense core MC35 by the IRAS deep survey \citep{Bei1992} means that there is no bright source with an upper limit of $\sim$0.1\,$L_{\odot}$. We checked the archival data of the {\rm Herschel} and {\rm Spitzer} and confirmed the non-detection of point sources at wavelengths shorter than 70\,$\mu$m. The derived dust temperature is as cold as 10--15\,K \citep{Hog2000} and the previous molecular gas survey in \jnco (1--0) with FCRAO could not find an outflow from this object \citep{Nar2012}. These observational signatures are consistent with the fact that there is no associated mature YSO. 
The single-dish observations found that this core has strong molecular line emission of high-density gas tracers, H$^{13}$CO$^+$, N$_2$H$^+$, and N$_2$D$^+$ \citep{Oni2002,Tob2013}, indicating that there is a density enhancement at the dust continuum peak. Figure \ref{fig:mc35} illustrates the observed area, the 1.2\,mm continuum emission by the IRAM 30\,m telescope, and infrared images by the {\rm Spitzer}. The observed central coordinate was ($\alpha_{J2000.0}$, $\delta_{J2000.0}$)=(4$^{\rm h}$35$^{\rm m}$7\fs5, +24\arcdeg09\arcmin17\farcs0).

\section{Results} \label{sec:results}

\subsection{Possible compact molecular outflow at the dust continuum peak} \label{result:outflow}

\begin{figure}[htbp]
\begin{center}
\includegraphics[width=180mm]{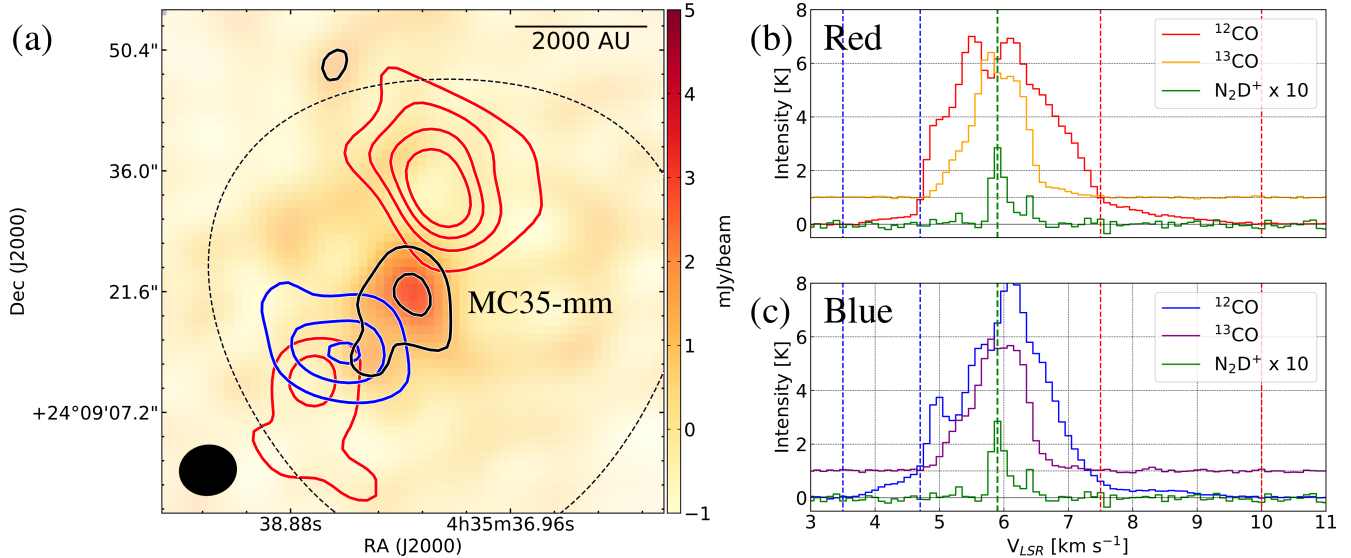}
\caption{(a) Integrated intensity distributions of the blueshifted and redshifted velocity components of the \jnco (2--1) emission are shown by the contours superposed on the 1.3\,mm image in the pseudo-color and black contours. The black lowest and subsequent contour levels are 3$\sigma$ and 6$\sigma$, respectively. The contour sequences of blue and red are [0.4, 0.6, 0.8] (K\,km\,s$^{-1}$) and [0.6, 0.8, 1.0, 1.2] (K\,km\,s$^{-1}$), respectively. The dashed black line indicates where the mosaic sensitivity falls to 50\%. Note that the primary beam attenuation is not corrected for the display purpose. The filled ellipse at the lower-left corner shows the synthesized beam of the ACA observation. (b) Average spectra of the redshifted component in \jnco(2--1) and \jsco(2--1) within the lowest red (north) contour showing in (a). The \jsco\ one is offset by +1\,K for the visualization.
The red and blue dashed lines represent the integrated velocity ranges of the contours in (a). The green solid and dashed lines show the \ntdp(3--2) profile toward MC35-mm and the systemic velocity of 5.9\,km\,s$^{-1}$ derived from the Gaussian fitting. (c) Same as panel (b), but for the blueshifted component.}
\label{fig:outflow}
\end{center}
\end{figure}

Figure\,\ref{fig:outflow} (a) shows the 1.3\,mm continuum emission toward MC35. We call the millimeter source $``$MC35-mm$"$ hereafter. Based on the two-dimensional Gaussian fitting to it, the central coordinate and deconvolved size are derived to be ($\alpha_{J2000.0}$, $\delta_{J2000.0}$) = (4$^{\rm h}$35$^{\rm m}$37\fs87, +24\arcdeg09\arcmin19\farcs8) and 10\farcs3 $\times$ 7\farcs2 with a position angle of $\sim$178\fdg1, respectively. The peak position coincides with that obtained by the IRAM 30\,m \citep{Mot2001}. This result means that the ACA observations traced the innermost part of the mass distribution. The peak and total continuum fluxes obtained by the ACA are 2.8\,mJy\,beam$^{-1}$ and 7.8\,mJy, respectively. Although the continuum flux is not the strongest one among our targets, we detect the strongest \ntdp\ emission with a peak brightness temperature of $\sim$0.5\,K at MC35-mm (Figure \ref{fig:outflow} (b), see also Sect. \ref{result:N2DP}).
This result means that there is a density enhancement at the position.

\begin{figure}[htbp]
\begin{center}
\includegraphics[width=180mm]{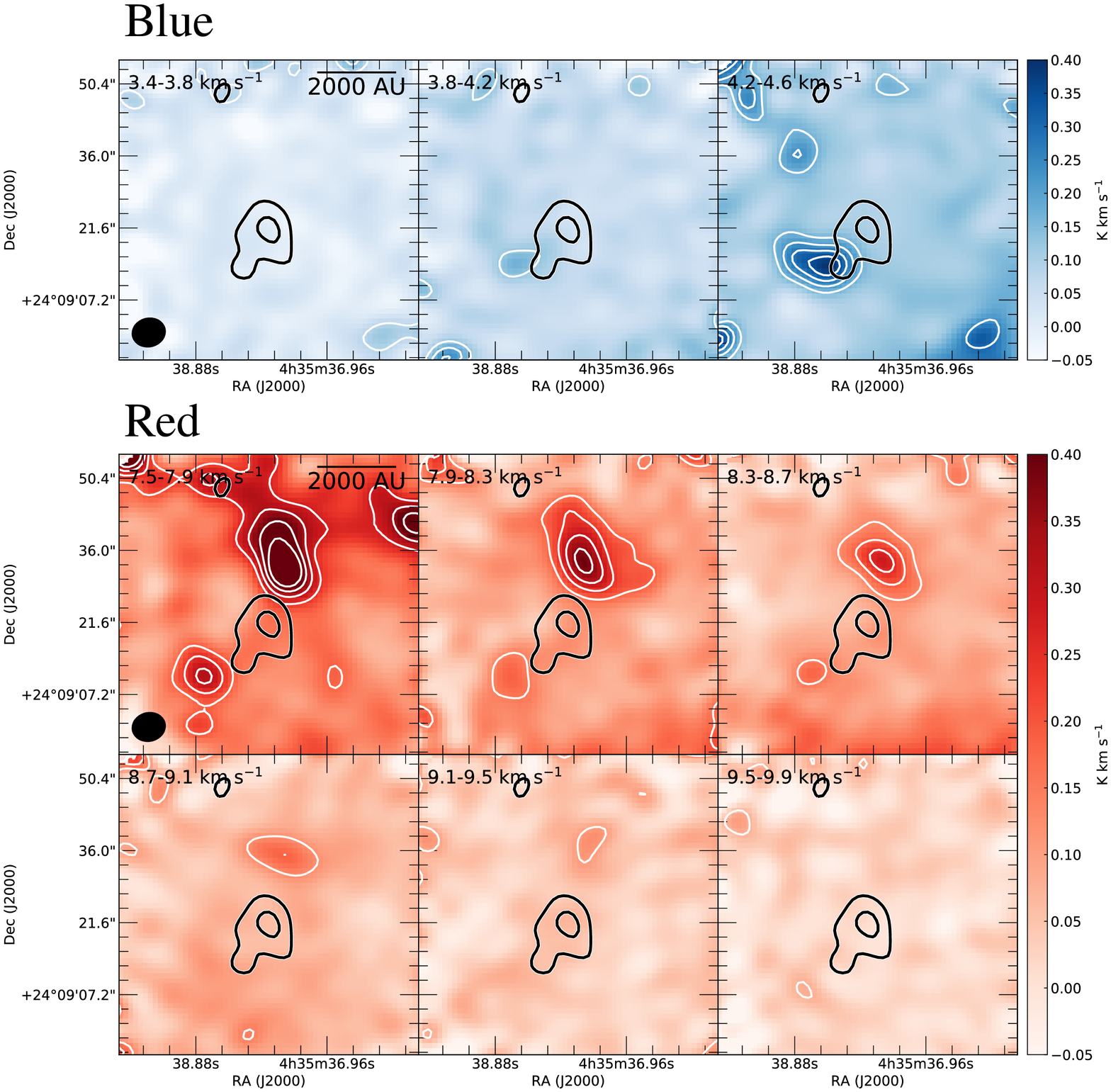}
\caption{Color-scale images and white contours show velocity-channel maps of the 7\,m + TP $^{12}$CO\,(2–-1) data toward MC35-mm. The lowest contour level and subsequent steps are 0.06\,K\,km\,s$^{-1}$. The black contours of each panel are the same as those in Figure\,\ref{fig:outflow} (a). The upper-left corner of each panel gives the integrated velocity ranges. The filled ellipses in the lower-left corner of the upper-left panel show the synthesized beam size.}
\label{fig:chmap}
\end{center}
\end{figure}

To find molecular outflows from deeply embedded objects, we investigated \jnco\ data toward the continuum-detected prestellar cores. Thus, we searched for high-velocity wing components more than a few km\,s$^{-1}$ apart from the systemic velocity as the excess emission to the Gaussian-like spectral profile by eye. As a result, we found that MC35-mm has indications of wing components. Figure\,\ref{fig:outflow} (a) shows the distributions of the high-velocity wing components in \jnco. As one can see, there are bipolar conical-shaped features of the blue/redshifted components, and they seem to be connected to the millimeter continuum peak. In the channel maps (Figure\,\ref{fig:chmap}), this nature is also well represented, especially at 4.2--4.6\,km\,s$^{-1}$ for the blue component and the 7.9--8.3\,km\,s$^{-1}$ for the red one.
Figure\,\ref{fig:outflow} (b) illustrates the velocity profiles of the \jnco\ emission. The selected velocity ranges of the blueshifted and redshifted components are 3.5--4.7\,km\,s$^{-1}$, and 7.5--10.0\,km\,s$^{-1}$, respectively. We could not detect significant $^{13}$CO emission in these velocity ranges. We can see clear line broadening up to $\sim$4\,km\,s$^{-1}$ with respect to the systemic velocity. Because the turbulence of this source ($\Delta V$ $\sim$0.3\,km\,s$^{-1}$, see \citealt{Tob2013}) cannot produce such a relatively high-velocity emission, there should be a driving source. The wing components are likely molecular outflow from a deeply embedded object inside MC35-mm. We found the blueshifted gas at the south side of the continuum peak, while there are two redshifted components at both the north and south sides. If the outflow is viewed away from pole-on, and it has a wide opening angle, the geometric effect can explain the present configuration. We could not find the blueshifted component at the northern side, possibly because the emission is very close to the systemic velocity with serious contaminations of the extended emission. 

We characterize the possible molecular outflow components. We defined the velocity difference between the central velocity of the \ntdp profile (5.9\,km\,s$^{-1}$) and the edges of the wing emission in \jnco\ (Figure \ref{fig:outflow}(b)) as its maximum velocities. The sizes of the wing components are the distances between the peak position of MC35-mm and the peak of the blue/red contours in Figure \ref{fig:outflow} (a).
Table\,\ref{table:outflow} summarizes the derived parameters from the observed values with an assumption of the inclination angles of the outflows as 30\arcdeg\ and 70\arcdeg with respect to the line of sight. The resultant dynamical time (=size/velocity) of the wing is $\sim$(1--6)\,$\times$\,10$^3$\,yr. We estimated the masses of the wing components assuming the local thermo-dynamical equilibrium with a uniform excitation temperature of 20\,K and [\jnco/H$_2$] = 10$^{-4}$, following the equations written by \cite{Pin2011}. We further discuss these parameters in Sect. \ref{firstcore}.

\begin{center}
\begin{table}[htbp]
\caption{Outflow properties}
\label{table:outflow}
\begin{tabular}{l|ccc|ccc|ccc} \hline
& \multicolumn{6}{c}{South}  & \multicolumn{3}{|c}{North}   \\ \hline
& \multicolumn{3}{c}{Blue} & \multicolumn{3}{|c|}{Red}  & \multicolumn{3}{c}{Red}   \\ \hline
Inclination angle (deg.)  & obs.  & 30   & 70   & obs.  & 30   & 70   & obs.  & 30   & 70     \\
Maximum velocity (km s$^{-1}$)  & 2.3  & 2.7  & 6.7  & 4.2  & 4.8  & 12.3  & 4.2  & 4.8  & 12.3    \\
Size (10$^3$\,au) & 1.5 & 3.0 & 1.6 & 2.1 & 4.2 & 2.2 & 1.6 & 3.1 & 1.6   \\
Dynamical time (10$^3$\,yr)   & $\cdots$ & 5.7 & 1.2 & $\cdots$ & 4.3 & 0.9 & $\cdots$ & 3.2 & 0.7   \\
Mass ($10^{-5}\,M_\odot$) & 0.6 & $\cdots$ & $\cdots$ & 1.2 & $\cdots$ & $\cdots$ & 2.0 & $\cdots$ & $\cdots$   \\
\hline
\end{tabular}
\end{table}
\end{center}

\subsection{Dense gas distributions traced by the \ntdp\ emission}\label{result:N2DP}

\begin{figure}[htbp]
\begin{center}
\includegraphics[width=180mm]{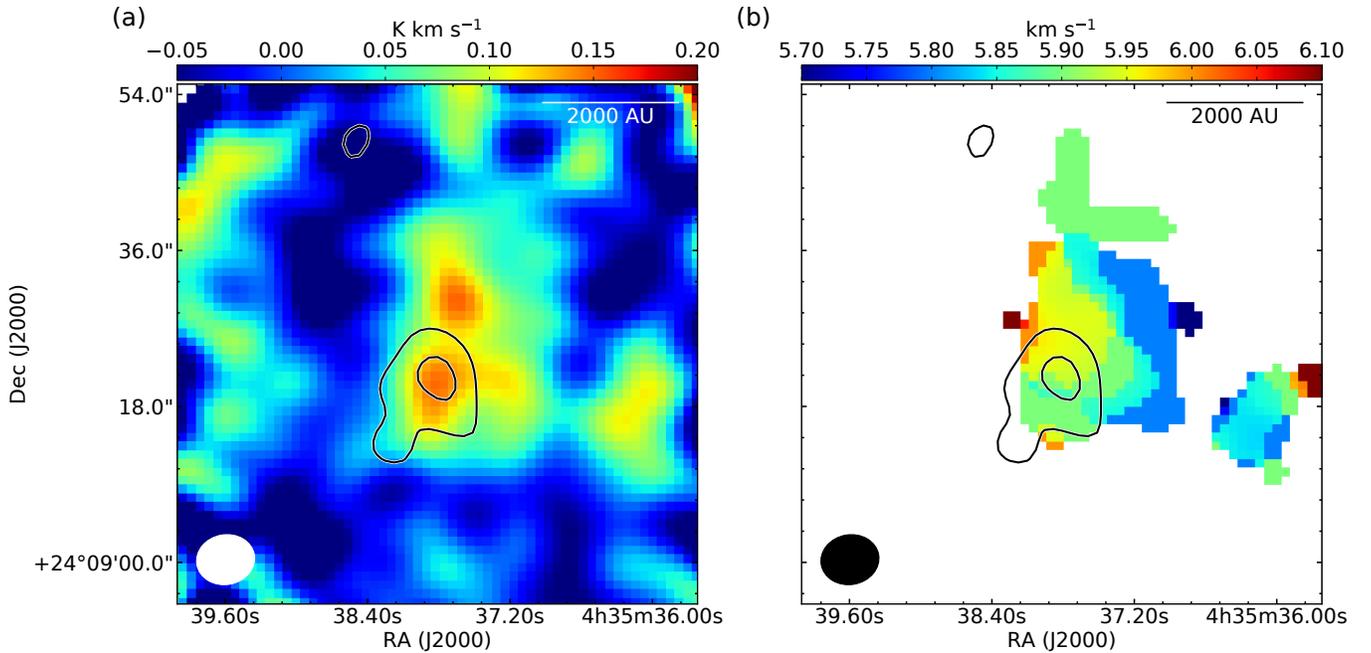}
\caption{ (a) The pseudo-color image shows the moment\,0 map of \ntdp(3--2) with the 7\,m array. The black contours in each panel are the same as those in Figure \ref{fig:outflow}. The filled ellipses at lower-left corner in each panel show the synthesized beam size. (b) The pseudo-color image shows the moment\,1 map of \ntdp(3--2) ($>$3$\sigma$).}
\label{fig:N2DP}
\end{center}
\end{figure}

Figure \ref{fig:N2DP} shows the distributions of \ntdp\,(3--2). The peak position of the velocity-integrated intensity (moment\,0) image corresponds to that in 1.3\,mm, indicating MC35-mm is in a cold/dense state, which is similar to evolved prestellar cores \cite[e.g.,][]{Cas2002}. The intensity-weighted mean velocity (moment\,1) map of \ntdp\ (Figure \ref{fig:N2DP} (b)) marginally shows a velocity gradient from the northeast to the southwest, which is roughly perpendicular to that of the wing components (Sect. \ref{result:outflow}). This feature may represent that there is a rotating component at MC35-mm.

\section{Discussions} \label{sec:dis}

\subsection{Is MC35-mm a candidate of the first core?} \label{firstcore}

We found possible outflow lobes at one of the densest $``$starless$"$ cores in Taurus. Before discussing the nature of the source, we mention the possibility of contamination emission by the nearby protostar, IRAS\,04325+2402. As shown in Figure \ref{fig:mc35}, there is a bipolar scattered-light nebula centered at the protostar. Based on the {\rm Spitzer} image, the dense region of MC35 seems to be illuminated by the scattered light. Although some previous studies reported a parsec-scale redshifted CO outflow from the protostellar system \citep[e.g.,][]{Hey1987,Nar2012}, its direction is inconsistent with that of the scattered-light bipolar nebula (see the red arrow in Figure\,\ref{fig:mc35}(b)). The Submillimeter Array (SMA) observation by \cite{Sch2010} has detected compact high-velocity CO components with a size of a few hundred au, whose elongation is roughly consistent with the scattered-light nebula. The scattered-light nebula traced by mid-infrared observations and molecular outflow distribution do not necessarily match completely, as discussed by \cite{Tob2010} and \cite{Tok2016}. Moreover, the bipolar feature of the newly detected wing components is hard to explain as an outflow lobe from another protostar. In summary, we conclude that the wing components are not just arising from the contamination of the protostellar activity in the southern part.

We estimate that the dynamical time of the outflow is approximately (1--6)\,$\times$\,10$^3$\,yr, depending on its inclination angle (see Sect. \ref{result:outflow}). This outflow is possibly the youngest one among other very low-luminosity objects (VeLLOs) or first-core candidates \citep{Dun2011}. The outflow velocity is not high compared to the typical low-mass protostar ($\gtrsim$10\,km\,s$^{-1}$, e.g., \citealt{Hog1998}), and the distribution shows a wide opening angle. These observational properties are consistent with that of another first-core candidate, L1451-mm \citep{Pin2011}, and the theoretical simulations, as mentioned in the introduction.  
The presence of the \ntdp\ emission (Sect. \ref{result:N2DP}) can be a key piece of evidence as a young phase. 
If there is a mature protostar, the \ntdp\ abundance rapidly decreases as the gas temperature increases. \cite{Tob2013} found two \ntdp\ peaks away from the protostar position and a drastic abundance decrease at the inner 1000\,au region in L1157. Unlike their observation, we do not find a clear position discrepancy between the \ntdp\ and continuum peaks. These features strongly indicate the absence of a mature protostar here, and thus we conclude that the MC35 core harbors a $``$candidate$"$ of the first core. The {\rm Herschel} Gould Belt survey \citep{Andre2014} detected the 100\,$\mu$m emission with an integrated luminosity of $\sim$0.6\,Jy, which is consistent with that of theoretical predictions \citep{Tom2010} and also comparable to the low-luminosity Class\,0 source \citep[e.g.,][]{Andre1999}. The only concern is that the 100\,$\mu$m peak does not completely match the ACA 1.3\,mm peak and is $\sim$16$\arcsec$ closer to the IRAS source.

We compare the evolutionary stage of MC35 with that of other condensations just before/after star formation in Taurus. 
IRAM\,041911+1522 is a low-luminosity ($\sim$0.08\,$L_{\odot}$) protostar \citep{Dun2006}. \cite{Andre1999} found the highly collimated CO outflow, and they concluded that the source is considered to be a very young accreting class\,0 protostar. Although MC27/L1521F was formerly categorized as a first-core candidate \citep{Miz1994,Oni1999}, {\rm Spitzer} observations revealed that the dense core harbors a protostar whose internal luminosity is $\sim$0.05\,$L_{\odot}$ \citep{Bou2006}. The subsequent ALMA studies found that the protostellar mass is $\sim$0.2\,$M_{\odot}$ without current mass accretion activity \citep{Tok2017}, and thus the protostar is more evolved than the first-core stage. L1544, which shares several similar properties with MC27/L1521F \citep{Cra2004}, is famous for a prototypical prestellar core. Although recent ALMA 12\,m array observations \citep{Cas2019} confirmed the centrally concentrated nature of the source, there is no report on the outflow so far. Among our starless sample in the ACA survey, MC5N has the strongest continuum emission (Paper~I), and we suggested that this core is a highly evolved stage on the verge of brown dwarf/very low-mass star formation \citep{Tok2019}. We could not detect the outflow toward MC5N as well, and the evolutionary stage is considered to be younger than MC35 and close to L1544. In summary, we suggest that MC35-mm is a unique source as the only first-core candidate in the Taurus region.

Note that it is observationally difficult to distinguish between the first core and VeLLO (see the review by \citealt{Dun2014}) so far; we need follow-up high-resolution studies to clarify the evolutionary stages, for example, by estimating the mass of the central object, as demonstrated by recent ALMA observations toward VeLLOs \citep{Tok2017,Lee2018}. 

\subsection{Lifetime of the first-core candidate} \label{lifetime}
The single-dish survey in Taurus found that there are $\sim$50 prestellar cores with an average density of $\sim$10$^5$\,cm$^{-3}$, and their lifetime is (4--5)\,$\times$\,10$^5$ yr (\citealt{Oni2002}; Paper~I). Because the theoretically predicted lifetime of the first core is typically $\sim$10$^3$\,yr \citep[e.g.,][]{Lar1969,Sai2006}, it is no wonder that no candidates are found within dozens of targets. However, our survey toward $\sim$30 prestellar cores in Taurus detected at least one promising candidate of the first core. This result observationally gives us a lifetime of the first core, $\sim$10$^4$\,yr, which is tentatively consistent with the dynamical time of the outflow lobes. Some theoretical studies suggested that the statistical lifetime of the first core can be longer than 10$^4$\,yr in the case of the gravitational collapse of a sub-solar-mass condensation \citep[e.g.,][]{Sai2006,Tom2010,Sta2018}. Although the core MC35 itself is as massive as $\sim$6--11\,$M_{\odot}$ \citep{Mot2001,Oni2002}, there is a possibility that the collapsing region is small. In fact, the parental core already forms the protostellar system, IRAS\,04325+2402, suggesting that the core fragmented into two pieces by some mechanism, and then the southern part collapsed locally. 

We discuss the possibility of a regional difference in the lifetime of the first core. Recent systematic investigations toward Perseus by \cite{Ste2019} using SMA reported that 6 of the 74 observed protostellar targets were suggested to be the first-core candidates. Note that not all of the 74 targets were confirmed to be protostars because they could not detect 1.3 and/or 0.85\,mm continuum emission toward several sources (see their Table\,7). They also discussed that at least two sources are relatively unlikely as the first-core candidates. The total number of the protostars can be revised into $\sim$70, and four of them are the first-core candidate. If we assume that the Class\,0/I phase lasts 0.5\,Myr \citep[e.g.,][]{Dun2014}, the inferred lifetime of the first-core (candidate) phase is $\sim$3\,$\times$\,10$^4$\,yr (=0.5\,Myr *4/70), which is significantly longer than the theoretically expected time. We thus speculate that some of the samples should contain more mature protostars or be in a prestellar phase. If only one source in Perseus is the first core, the inferred lifetime is less than 10$^{4}$\,yr, which is consistent with that of our Taurus result. As suggested by \cite{Ste2019}, L1451-mm is the best candidate in Perseus.

If the Perseus candidates are all genuinely a first core, the environment may play a vital role in extending the first-core phase. In general, low-mass cluster-forming regions show fragmented core distributions and highly turbulent gas kinematics (see \citealt{Tac2002,War2007}) compared to those in isolated star-forming region, like Taurus. From a theoretical perspective, rotation of a parental core can work to extend the first-core phase \citep{Sai2006}. If the more turbulent environment forms a larger number of sub-solar-mass cores \citep{Enoch2006} with rapid rotation, this can explain the lifetime difference in the first-core phase between the two regions. 
However, our current observational understanding regarding the first-core candidates is still incomplete. Detailed follow-up observations using millimeter/submillimeter interferometric facilities will allow us to understand the true nature of such candidates further. 

\acknowledgments
This Letter makes use of the following ALMA data: ADS/ JAO.ALMA \#2018.1.00756.S. ALMA is a partnership of ESO (representing its member states), NSF (USA) and NINS (Japan), together with NRC (Canada), MOST and ASIAA (Taiwan), and KASI (Republic of Korea), in cooperation with the Republic of Chile. The Joint ALMA Observatory is operated by ESO, AUI/NRAO, and NAOJ. This work was supported by NAOJ ALMA Scientific Research grant No. 2016-03B and Grants-in-Aid for Scientific Research (KAKENHI) of Japan Society for the Promotion of Science (JSPS; grant Nos. 18K13582, and 18H05440). 

\bibliographystyle{aasjournal}

\end{document}